%% file: cas-refs.tex
\documentclass[lettersize,journal]{IEEEtran}
\usepackage{amsmath,amsfonts}
\usepackage{algorithmic}
\usepackage{algorithm}
\usepackage{array}
\usepackage[caption=false,font=normalsize,labelfont=sf,textfont=sf]{subfig}
\usepackage{textcomp}
\usepackage{stfloats}
\usepackage{url}
\usepackage{verbatim}
\usepackage{graphicx}
\usepackage{cite}
\usepackage[square,numbers]{natbib}
\usepackage{comment}
\usepackage{listings}
\usepackage[section]{placeins}
\usepackage{tabularx}
\usepackage{xcolor}

\newcolumntype{b}{X}
\newcolumntype{s}{>{\hsize=.5\hsize}X}
\newcolumntype{t}{>{\hsize=.2\hsize}X}

\begin{document}

\title{An IoT Architecture Leveraging Digital Twins: Compromised Node Detection Scenario}

\author{\IEEEauthorblockN{Khaled Alanezi}\\
\IEEEauthorblockA{\textit{Department of Computing,}
\textit{College of Basic Education, PAAET, Kuwait}\\
kaa.alanezi@paaet.edu.kw}\\
\and
\IEEEauthorblockN{Shivakant Mishra}\\
\IEEEauthorblockA{\textit{Computer Science Department,}
\textit{University of Colorado, Boulder, USA}\\
mishras@colorado.edu}

\thanks{\textbf{This work has been submitted to the IEEE for possible publication. Copyright may be transferred without notice, after which this version may no longer be accessible.}}}

\markboth{Journal of \LaTeX\ Class Files,~Vol.~14, No.~8, August~2021}%
{Shell \MakeLowercase{\textit{et al.}}: A Sample Article Using IEEEtran.cls for IEEE Journals}


\maketitle

\begin{abstract}
Modern IoT (Internet of Things) environments with thousands of low-end and diverse IoT nodes with complex interactions among them and often deployed in remote and/or wild locations present some unique challenges that make traditional node compromise detection services less effective. This paper presents the design, implementation and evaluation of a fog-based architecture that utilizes the concept of a digital-twin to detect compromised IoT nodes exhibiting malicious behaviors by either producing erroneous data and/or being used to launch network intrusion attacks to hijack other nodes eventually causing service disruption. By defining a digital twin of an IoT infrastructure at a fog server, the architecture is focused on monitoring relevant information to save energy and storage space. The paper presents a prototype implementation for the architecture utilizing malicious behavior datasets to perform misbehaving node classification. An extensive accuracy and system performance evaluation was conducted based on this prototype. Results show good accuracy and negligible overhead especially when employing deep learning techniques such as MLP (multilayer perceptron). 
\end{abstract}

\begin{IEEEkeywords}
Internet of Things, Digital-Twin, Fog Computing, Compromised Node Detection.
\end{IEEEkeywords}

\input{Introduction}
\input{Design.tex}
\input{Related.tex}
\input{Experimental.tex}
\input{Evaluation.tex}
\input{Conclusion}

\bibliographystyle{IEEEtran}
\bibliography{cas-refs.bib}

 




\vfill

\end{document}

%% file: Introduction.tex
\section{Introduction}
\label{sec:introduction}

The Internet of Things (IoT) has brought increased convenience and productivity to homes, factories, malls, hospitals, sidewalks, city squares and more. Typically, an IoT application is a distributed system with components deployed across the IoT-Fog-Cloud continuum. IoT components perform data collection from the environment and feed the data to smart decision-making systems running in the cloud with the fog layer playing a staging role for data filtering and aggregation. Building applications on top of such complex infrastructure is inherently difficult thereby leading to interoperability issues \cite{noura2019interoperability}. 
Digital-Twins (henceforth DT) has been proposed as a tool for managing these systems, predicting their behavior under different scenarios and improving their performance. DT is defined as a virtual replica of a physical entity such as people, assets, systems or processes \cite{jeong2022digital}. The concept of DT is not new, however, as it stems from existing mature concepts such as IoT, AI and big data \cite{raes2021duet}. Despite showing clear benefits for IoT, a systems implementation showing how to utilize DTs in an IoT application and how the interplay of different components affects the performance is still lacking. This work aims at filling this gap while focusing on compromised node as an example scenario. 

While IoT applications have enhanced our lifestyles in many areas, there is an associated cost of potentially exposing user's privacy and security. Typically, IoT nodes embed wide array of sensors and are connected to the network, thereby making them prime targets for attackers. Once an IoT node is exploited, an attacker would typically couple the attack with payload codes that can achieve certain objectives. First, an attacker can carry a Cyber-Physical attack \cite{wu2019detecting} to introduce \textbf{data anomalies} causing control system failure thereby leading to physical damage  \cite{stuxnet,steel-mill}. Second, once inside the network (i.e. \textbf{network intrusion}), an attacker can utilize the node to launch large scale botnet attacks such as the Mirai IoT Botnet \cite{kumar2020early}) which can lead to large scale service disruption.

While there has been a plethora of research done in the area of preventing and/or detecting node compromises over the last 30+ years, compromised node detecting in an IoT environment presents some unique challenges. First, typically only weak security measures are taken to protect IoT nodes, which can be attributed to the focus on keeping the node cost low and ensuring a plug-and-play operation, so that they can be deployed in large numbers. Second, these nodes are often placed remotely in the wild with 
little or no monitoring with many IoT vendors failing to provide proper measures or automated-tools for timely security updates \cite{neshenko2019demystifying}.
Third, the sheer number and variety of IoT devices in an IoT infrastructure typically running into hundreds and even thousands means an increased attack surface available to the attackers and for the system administrators to monitor and maintain. Finally, complex interplay among thousands of nodes makes it difficult to identify faulty or anomalous behavior often confined to a small number of nodes under a variety of different scenarios.

To address these challenges, we propose a fog-based architecture that utilizes the concept of a {\it digital-twin} in which a virtual replica of the IoT infrastructure is created in the fog to momentarily replicate IoT node state changes where threat detection and mitigation can take place. Our proposed architecture uses DT to employ detection techniques for both data anomaly and network intrusion attacks for enhanced IoT network security. Our architecture benefits from using DTs by utilizing two of its conceptual features \cite{minerva2020digital}. First, by definition, the DT concept proposes to only mirror properties and characteristics that are of importance to the application context. Consequently, we only reflect status information represented by sensor values and network activity summary information in the DT. Clearly, by focusing only on important information our solution can save device energy, fog server space and network resources. Second, creating DTs for all IoT nodes paves the way for understanding the aggregate behavior of those nodes thereby leading to better governance and control of the complex IoT system. In summary, the contributions of this work are three fold:
\begin{enumerate}
    \item We provide a detailed design of a fog-based architecture utilizing the concept of DTs for compromised node detection in an IoT environment.
    \item We show how this architecture can utilize various models for compromised IoT node based on data anomaly and network intrusion detection. We also report the accuracy and performance of those models.
    \item We provide a prototype implementation and extensive performance evaluation of the architecture by utilizing open-source solutions such as Ditto \cite{ditto} and Docker \cite{docker}.
\end{enumerate}
The flow of the remaining sections of the paper is as follows. In Section \ref{sec:design} we present the design of the architecture and introduce the details of the datasets utilized to build the data anomaly and network intrusion models. After that, Section \ref{sec:related} presents related works followed by Section \ref{sec:experimental} which provides technical details of the prototype implementation for the architecture and the models. In Section \ref{sec:evaluation} we present system performance measurements for the prototype before concluding in Section \ref{sec:conclusion}.

%% file: Design.tex
\section{Design}
\label{sec:design}
\subsection{Architecture}
\label{subsec:overall}
\begin{figure*}
    \centering
\includegraphics[width=0.6\textwidth]{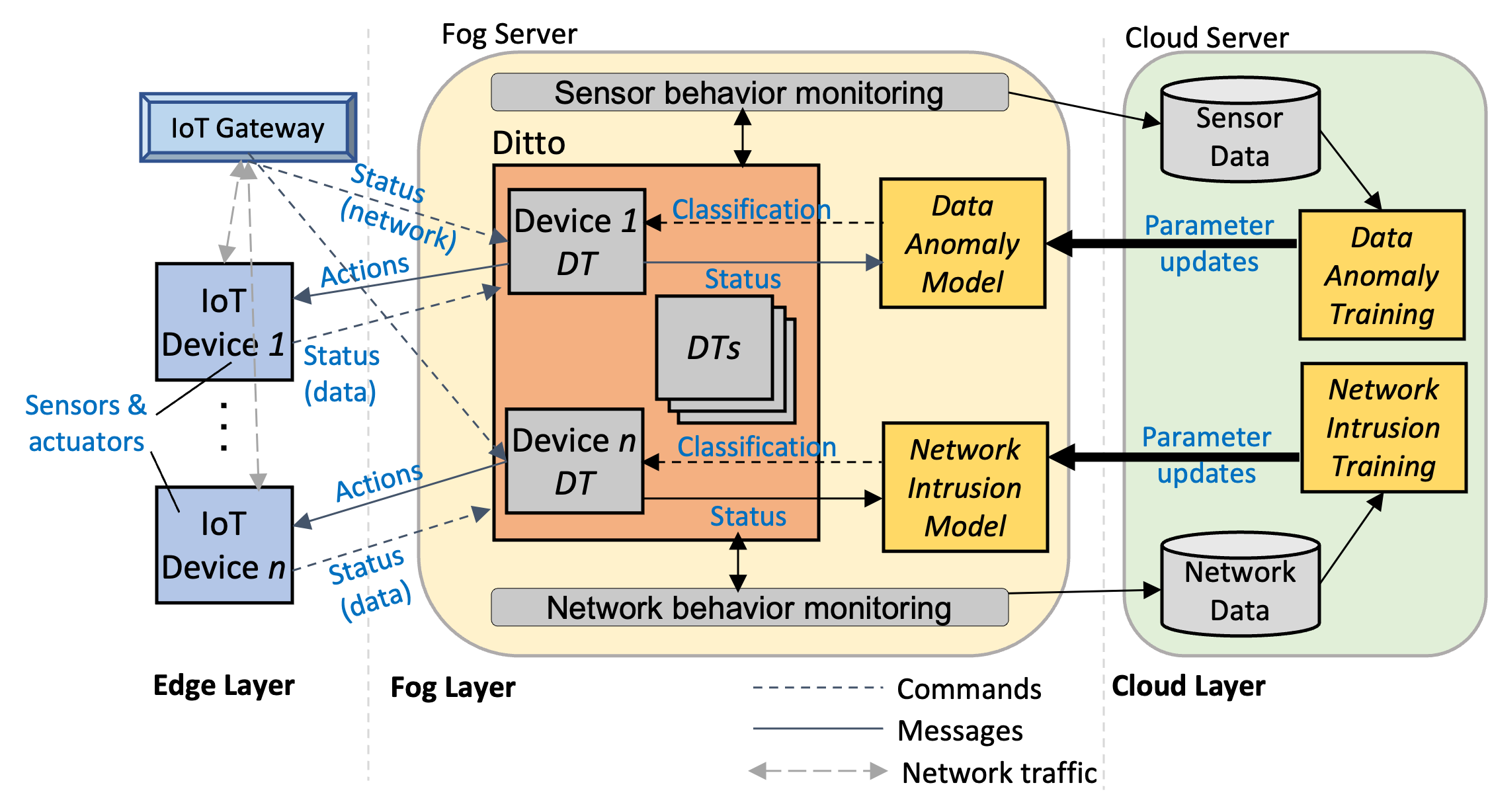}
    \caption{Solution Architecture}
    \label{fig:architecture}
\end{figure*}
As shown in Figure \ref{fig:architecture}, the proposed architecture spans the edge-fog-cloud continuum. The \textbf{edge layer} typically consists of battery-powered \textbf{IoT devices} with limited resources. Those devices are meant to perform sensing and actuation in the environment. At the other extreme end of the architecture is the \textbf{cloud layer}, which typically has abundant storage and computing resources. The architecture utilizes a \textbf{cloud server} to perform compute-intensive training of malicious node detection models. In the middle of the architecture is the \textbf{fog layer} providing low-latency access for IoT devices to the \textbf{fog server}. We propose to utilize an edge server \cite{satyanarayanan2017emergence} installed in the fog to provide low-latency access to trained malicious node detection models. These models will be harnessed to monitor nodes activity at the edge of the network to detect malicious behavior on spot. 

We now turn into describing the functionality of each component in the architecture. Starting from the edge layer, the \textbf{IoT devices} have the sensing and actuation capabilities to aid various IoT applications. Those devices are connected to the network using an \textbf{IoT gateway}. Devices at the edge layer have one-network-hop access to a \textbf{fog server} running Ditto \cite{ditto}. Ditto is an open source framework for managing DTs. Through APIs exposed by Ditto, \textbf{device DTs} receive status updates from IoT devices reflecting corresponding device status change (i.e. sensor data changes). In addition, DTs receive status update from the \textbf{IoT gateway} that are related to the network behavior of the device (Detailed design of DTs is discussed in Section \ref{subsec:dt-design}). 

DT status updates received from the IoT environment are used in two ways. First, the sensory status and the network status are used as input for the \textbf{data anomaly model} and \textbf{network intrusion model} respectively to discover any anomalies in node behavior. The classification result is sent to the corresponding DT, and only if the result shows that there is a possible anomaly, a message is sent from the DT to the corresponding device to take the needed action. Note that the actions themselves are defined by the system administrators, e.g. quarantine the device or shut it down to avoid infecting other nodes in the environment. The benefit from combining results from both classifiers will lead to decreased false negatives thereby increasing confidence in the overall system. The second usage of reported status updates is to forward them via the \textbf{Sensor/Network behavior monitoring} components to the \textbf{Sensor/Network databases} in the cloud. The latter will act as the ground truth for the \textbf{Data anomaly} and the \textbf{Network intrusion} training done periodically. Note that behavior monitoring should involve system administrators utilizing malware analysis techniques for detecting IoT malware \cite{liu2019integrated}. Once a model is updated in the cloud by retraining, new parameters of the model are pushed to the fog server to be used. The periodic updates ensure that the architecture is adaptive to changing norms in the environment. We discuss the design of the data anomaly model and network intrusion models in sections \ref{subsec:data-anomaly-model} and \ref{subsec:network-intrusion-model} respectively.

\subsection{Digital Twin Design}
\label{subsec:dt-design}
\begin{figure}
    \centering
\includegraphics[width=0.35\textwidth]{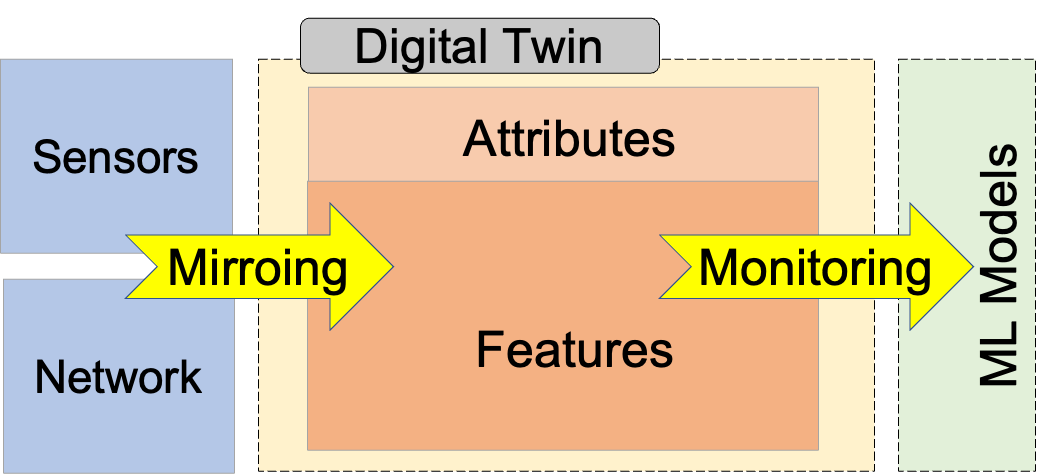}
    \caption{Digital Twin Deployment Status}
    \label{fig:digital-twin-design}
\end{figure}
The concept of DTs has been around for almost two decades now. However, a standard design and implementation guidelines for the concept is still lacking \cite{schroeder2020methodology}. Consequently, the design approach for the DT in this project will follow the 5-steps guided approach for DT evolution \cite{jeong2022digital} of mirroring, monitoring, simulation, federation and autonomous. We begin by covering the first two steps of mirroring and monitoring. The remaining three steps represent later design stages that build on the understanding of the first two stages, which we leave as a future work as described in Section \ref{sec:conclusion}. 
\begin{table*}[htbp]
\caption{Utilized Datasets for Data Anomaly and Network Intrusion Models}\label{table:datasets-summary}
    \centering
    \begin{tabularx}{\textwidth}{ssssbb}
        \hline
        Dataset Name &Scenario &No. of Records &No. of Features &Sample Features     &Labels\\ 
        \hline
        AnoML-IoT \cite{kayan2021anoml} & Data Anomaly &6,558 &6 &Timestamp, temperature reading in C and humidity percentage. &Normal and Anomalous \\ 
        DS2OS \cite{aubet2018ds2os} &Data Anomaly &357,941 &13 &Timestamp, src ID, src type, dest. type, operation and value. & Normal and Anomalous (Probing, DoS, Mal. Control, Mal. Operation, Scan, Spying and Wrong Setup.) \\ 
        IoTID20 \cite{ullah2020scheme} &Network Intrusion &625,783 &86 &Timestamp, flow duration, min/max fwd packet size and flow bytes/s. &Benign and Malicious (Mirai, DoS, Scan and MITM ARP Spoofing.) \\ 
        \hline        
    \end{tabularx}
\end{table*}
As can be seen in Figure \ref{fig:digital-twin-design}, the are two types of information mirrored about an \textbf{IoT device} in the DT. First, the \textbf{attributes} of the device cover static metadata such as the unique identifier of the device, the manufacturer and the location coordinates if any. Second, the \textbf{features} contain dynamic information that changes over time and are replicated to the DT momentarily. For data anomaly monitoring we mirror changing sensors' values as captured by the IoT node micro-controller and the calls that the IoT node make to other nodes or services. On the other hand, for network monitoring we mirror summary information related to the exchanged network packets by the device. The exact sensor values and network summary status that must be replicated to the DT depends on the input parameters of the utilized data anomaly and network intrusion models. Sections \ref{subsec:data-anomaly-model} and \ref{subsec:network-intrusion-model} describe three machine-learning models that we will evaluate for use in the architecture. Whenever any of the models is actually deployed on the fog server, its corresponding features must be replicated to the DT in order to be used by the machine-learning model for device behavior classification. We evaluate the performance of the models when integrated with the DT in Section \ref{sec:experimental}.    
\subsection{Data Anomaly Models}
\label{subsec:data-anomaly-model}
Monitoring data anomaly involves inspecting data generated from sensors to be consumed by an application or a service. Table \ref{table:datasets-summary} lists three example datasets we used for the compromised node detection along with summary information for each dataset. First, the \textbf{AnoML-IoT} dataset contains sensor readings captured over two days for temperature, humidity, light, loudness and air quality sensors. The authors of this dataset created data anomalies by subjecting the sensors to an air dryer for some periods of time. Data coming during this time is marked as an anomaly. Otherwise, sensor readings are marked as benign. Second, \textbf{DS2OS} is a synthetic dataset created by the Distributed Smart Space Orchestration System. The data is generated from a simulated setup including four IoT sites each containing various services that can call each other. Example services include thermostat and door lock controllers. A service can read or write a value to or from another service depending on the required action. Beside the normal traffic, the authors simulated communications for various attacks including data probing, denial-of-service, malicious control, malicious operation, scan, spying and wrong setup. In summary, we will use the first two datasets as an example for an IoT deployment generating malicious behavior at the perception layer (i.e. data anomalies).

\subsection{Network Intrusion Models}
\label{subsec:network-intrusion-model}
In addition to false data injection attacks to confuse an application, a compromised node could be exploited by an attacker to launch network attacks such as port scans, DDoS attacks or spoofing. These attacks often lead to data theft, data corruption or system failures. In this part of the architecture we focus on discovering and stopping network intrusion attacks. In order to achieve this goal, we utilized the \textbf{IoTID20} dataset also described in Table \ref{table:datasets-summary}. This dataset is built using an actual IoT network consisting of two smart home devices (AI Speaker and Security camera) and multiple laptops and smartphones connected to an isolated LAN. Various attacks are then deployed or simulated using this setup such as Mirai, Denial-of-Service (DoS), Scan and Man-In-The-Middle Address Resolution Protocol (MITM ARP) spoofing attacks. Network packet captures are then extracted and network packets containing summary features of communication sessions are labeled as benign or malicious accordingly.

%% file: Related.tex
\section{Related Work}
\label{sec:related}
This work investigates \textbf{integrating DTs in IoT systems}. The application served by the resultant architecture is compromised node detection by utilizing  \textbf{data anomaly} and  \textbf{network intrusion}. We summarize in this section the recent works in each of these areas.
\subsection{Digital Twins in IoT}
Digital-Twins DTs, a concept popular in manufacturing, has been proposed mainly for use in Industrial IoT (IIoT). With the aid of DT concept, an IIoT controller collects massive data from devices to take smart environment-wide decisions \cite{jiang2021digital,canedo2016industrial}. To serve this purpose, a reference model of DTs in IIoT is proposed \cite{delfino2019industrial} which involves mirroring the internal structure, runtime environment, APIs of the physical objects as well as features such as scalability, interoperability, security and privacy. 

Also, the concept of DTs has been recently proposed in the IoT context for staff safety management in cold warehouses' hazardous environments \cite{zhao2021iot}. Here, a DT is created to synchronize the time/space information of the staff with the controller. Consequently, the controller will run algorithms to detect staff motionless status so as to alert first aid workers in a timely manner. In addition to safety, the concept of DTs is utilized in IoT context for elderly health \cite{liu2019novel} where DTs are used to mirror medical data obtained from wearable devices to monitor and diagnose health issues.

Furthermore, the concept of DTs is utilized to enable smart buildings and smart cities. For the former, DTs replicate static and dynamic building data for enhanced building monitoring and management \cite{eneyew2022toward}. Whereas for the latter, DUET \cite{raes2021duet} is proposed as a DT framework built for smart city applications. DUET creates a cloud of integrated DT models that can be queried to perform smart city planning (i.e. traffic, air pollution, or noise pollution) at the city level.

DTs are also proposed in IoT for smart agriculture systems \cite{verdouw2021digital} where the goal is to monitor farm information remotely thereby reducing manual efforts and to simulate the effect of intervention techniques on farm productivity. Also, battery energy storage systems BESS \cite{kharlamova2023evaluating} are also shown to benefit from the DT concept so as to prevent possible failures and cyberattacks.

Lastly, a survey paper by Minerva et al. \cite{minerva2020digital} discussed DTs features and architectures in the context of IoT. 

\subsection{Data Anomaly Detection in IoT}
IoT applications involve data collection to provide controlled services to users. Consequently, an attacker can impact the accuracy of the controller creating anomalies in sensory data \cite{khan2019malicious}. There are several works aimed at detection of data anomaly in IoT. AnoML-IoT \cite{kayan2021anoml} presents a pipeline for data anomaly detection with the goal of masking heterogeneity in IoT systems. On the other hand, a preliminary work by Hussain Et al. \cite{hussain2022explainable} focuses on utilizing explainable AI (XAI) to understand the output of deep learning data anomaly techniques for IoT. Lastly, a survey paper \cite{erhan2021smart} presented a summary of techniques to detect data anomalies in IoT sensory systems and the impact of Cloud-Fog-Edge architectures on them. 

\subsection{Network Intrusion Detection in IoT}
Malicious node behavior other than producing data anomaly could manifest in performing anomalous network activity. Researchers created datasets \cite{ullah2020scheme,iot-23,liu2020machine} containing the network traffic of benign and malicious nodes in order to build models that can automatically detect malicious traffic. Based on such datasets various techniques and architectures to detect malicious network behavior are proposed. EDIMA \cite{kumar2020early} is a full architecture geared towards detection of network intrusion activity during the scanning phase instead of the attack phase. Diro Et Al. \cite{diro2018distributed} proposed the use of deep learning models over tradition ML in order to cope with newly surfacing small mutations of malware. Also, Convolutional Neural Networks CNN were proposed \cite{ullah2021design} for malicious traffic detection and classification. Finally, running network traffic classification models on resource-limited IoT microcontrollers is also investigated \cite{dartel2021malware}.

%% file: Experimental.tex
\section{Experimental Work and Implementation}
\label{sec:experimental}
We propose a dynamic architecture to detect malicious IoT nodes. This requires real-time monitoring of node behaviors and flagging of observed anomalies as they occur. To achieve this objective, the architecture must support two requirements. First, the ability to capture relevant node behavior. Second, the availability of dynamically-trained machine-learning models for online behavior classification. We begin this section by describing the hardware and software components involved in building the architecture and how these components are integrated. These details are described in Section \ref{subsec:architecture-impl}. After that, we describe the classification models and their accuracy based on various machine-learning classifiers when applied to the chosen datasets. As stated in Section \ref{sec:design}, we propose that the architecture must train two types of models namely the data anomaly model and the network intrusion model. We report in this section the design and accuracy for the two models when trained with various classifiers. Note that the models along with the codes to build them are made publicly available on OSF\footnote{\url{https://osf.io/mh6es/?view_only=f7dce520b1b64ce198ab039563c29e5f}}.
\subsection{Architecture Implementation}
\label{subsec:architecture-impl}
\begin{table*}[htbp]
\caption{Hardware and Software Components Used to Build the Architecture Prototype.}\label{table:prototype-components}
    \centering
    \begin{tabularx}{\textwidth}{stb}
        \hline
        Component Name & Type & Role \\ 
        \hline
        MacBook Air & Hardware & Fog server. \\
        Arduino Uno R3 \cite{uno} & Hardware & IoT device Microcontroller. \\
        Arduino Ethernet Shield \cite{ethernet-shield} & Hardware & Connects microcontroller to the LAN. \\
        Huawei Wireless Router & Hardware & Wireless LAN router.\\
        Ditto \cite{ditto} & Software & A framework for implementing Digital Twin Capabilities for IoT. \\
        Docker \cite{docker} & Software & Container engine for building Ditto, data anomaly \& network intrusion containers. \\
        Data Anomaly Model Container & Software & Container running the data anomaly model. \\
        Network Intrusion Model Container & Software & Container running the network intrusion model. \\
        Postman \cite{postman} & Software & Software platform used to call Ditto to create digital twins. \\
        \hline        
    \end{tabularx}
\end{table*}

We built a prototype for the architecture presented in Section \ref{sec:design}. The prototype is a distributed solution spanning the edge of the network and the fog layer as described earlier. The components utilized to build this prototype are listed in Table \ref{table:prototype-components}. Note that the code for this prototype is also shared under the same OSF project below the \textbf{Online Experiment} component.

We utilized a MacBook Air laptop to resemble a fog server. The fog server is typically a tethered (powered) machine with good computing capabilities that is one-network-hop away from edge nodes. On the other hand, an Arduino Uno R3 board was used to resemble the IoT node. We stacked an Ethernet shield with the Arduino Uno to provide it with Ethernet capability and connected it along with the edge server to the same LAN using a Huawei wireless router. 

The prototype relied on Eclipse Ditto \cite{ditto} for implementing the needed DT functionality. This functionality includes configuring DTs, online replication of DT state from the physical IoT node to the virtualized replica and observing DT state changes. To run Ditto, we deployed the pre-built ditto docker images on the edge server. 

We also used Docker \cite{docker} to dockerize and deploy python scripts for loading the machine-learning models used for performing online classification. Each script will load a pre-trained model that is stored on disk in a pickle file (*.pkl) \cite{pickle} and listen to a TCP socket to receive requests for record classification. We evaluated the performance of the overall architecture using the built prototype in Section \ref{subsec: arch-performance-eval}.

DTs must be configured on Ditto before the mirroring and monitoring processes take place. Listing \ref{listing:DT-config} demonstrates an example JSON configuration for creating a new DT in Ditto. This configuration pertains to the case of monitoring data anomalies based on the knowledge gained from the AnoML-IoT dataset as described in Section \ref{subsec:data-anomaly-model}. Note that we only cover the AnoML-IoT scenario in our online implementation which is sufficient to measure the system performance of the architecture. Ditto requires to identify a \textbf{definition} clause, which should contain a unique identifier for the DT. This identifier will be used for later communication with the DT to update/observe its state features values as they change. The \textbf{attributes} section of the JSON document contains static information for the DT such as its logical location, manufacturer, model and so on. Finally, the \textbf{features} section tracks the online status of the DT. In the case of the AnoML-IoT scenario shown, it lists the readings of the four sensors as captured by the IoT node that will be monitored to detect anomalies in the environment. 

Section \ref{subsec: arch-performance-eval} evaluates the systems performance of the architecture while utilizing various datasets/ML classifiers combinations.
\begin{lstlisting}[frame=single,basicstyle=\small,caption=DT Configuration\label{listing:DT-config}]
{
 "definition":"kw.edu.paaet:arduino:1.0",
 "attributes":{
  "manufacturer":"Arduino Inc",
  "location":"CS Dept. Corridor",
  "serialno":"1",
  "model":"Arduino Uno"
 },
 "features":{
  "temperature":{"properties":{"value": 0.0}},
  "humidity":{"properties":{"value": 0.0}},
  "light":{"properties":{"value": 0.0}},
  "loudness":{"properties":{"value": 0.0}}
 }
}
\end{lstlisting}
\subsection{Data Anomaly Models Implementation}
\label{subsec:data-anomlay-accuracy}
We utilized two datasets for data anomaly detection. First, the AnoML-IoT dataset represents a scenario where the attacker is deliberately producing erroneous sensor values or data anomalies. Second, the DS2OS also included an attacker controlling the application as it contains data traces produced at the application level. However, instead of producing erroneous sensor values, in this dataset the application is producing malicious calls for other application level services in the same environment.
For each dataset we ran three types of classifiers namely Random Forest RF, Support Vector Machines SVM and Multilayer Perceptron MLP. The accuracy results are shown in Table \ref{table:data-anomaly-accuracy}. Random forest was chosen due to its ability of visualising and studying the classification results. We kept the number of estimators to 100, the default number in Scikit Learn \cite{Scikit}. We also used SVM due to its efficiency with numerical as well as categorical features since the datasets contained both types of features. Finally, we used MLP as we wanted to also include a deep learning classifier in the experimental design. MLP is known to perform well with tabular data similar to the datasets we use. We have chosen to implement MLP with 3 hidden layers and 11 neurons at each layer as a starting point. We noticed that we were able to train the datasets efficiently with this choice. Optimizing MLP is an iterative process that is out of the scope of our work since our focus is to measure the system performance of the architecture.
\begin{table*}
  \centering
  \caption{Accuracy for Data Anomaly Classifiers Based on Random Forest (RF), Support Vector Machines (SVM) and Multilayer Perceptron (MLP)}
  \label{table:data-anomaly-accuracy}
  \begin{tabular}{lcccccc}
    \hline
     & \multicolumn{3}{c}{AnoML-IoT} & \multicolumn{3}{c}{DS2OS} \\
     \cline{2-7}
     & RF(n=100) & SVM & MLP & RF(n=100) & SVM & MLP \\
    \hline
    Accuracy & 0.989 & 0.938 & 0.966 & 0.993 & 0.993 & 0.993 \\
    Precision & 0.989 & 0.952 & 0.977 & 0.986 & 0.984 & 0.984 \\
    Recall & 0.983 & 0.899 & 0.942 & 0.893 & 0.887 & 0.894 \\
    F1-Score & 0.986 & 0.921 & 0.958 & 0.934 & 0.930 & 0.934 \\
    \hline
  \end{tabular}
\end{table*}
We now turn into comparing the accuracy results for the three classifiers within each dataset as seen in Table \ref{table:data-anomaly-accuracy}. For the AnoML-IoT dataset, a relatively small dataset, we noticed that RF produced the best results in terms of the four metrics (i.e. Accuracy, Precision, Recall and F1-Score). However, the difference margin is small since for example the largest gap is between RF and SVM at the recall metric. Recall reflects the ability of the classifier to correctly flag all cases positive and negative. Note that this measure was the most difficult for all the three classifiers.

For the DS2OS dataset, still Table \ref{table:data-anomaly-accuracy}, a dataset that is larger than the AnoML-IoT (refer to Section \ref{subsec:data-anomaly-model} for datasets description), all the three classifiers produced the same accuracy of 99.3\%. The accuracy is a common metric that reflects the ability of the classifier to correctly flag positive and negative instances. Also in this dataset, the recall achieved the worst results with RF producing the lowest result. Conversely, RF achieved the best result when it comes to precision. 

\subsection{Network Intrusion Models Implementation}
\label{subsec:nw-intrusion-accuracy}
For the network intrusion, we utilized the IoTID20 dataset which consists of features extracted from packet capture (*.pcap) files containing packet captures for attack scenarios as well as well as benign scenarios. This dataset is a typical dataset to be used in intrusion detection solutions. Detecting malicious network behavior for an IoT node requires capturing the network packets produced by this IoT node and extracting features from those packets to be used by the classifier.

\begin{table*}
  \centering
  \caption{Accuracy for Network Intrusion Classifiers Based on Random Forest (RF), Support Vector Machines (SVM) and Multilayer Perceptron (MLP)}
  \label{table:network-instrusion-accuracy}
  \begin{tabular}{lccc}
    \hline
     & \multicolumn{3}{c}{IoTID20} \\
     \cline{2-4}
     & RF(n=100) & SVM & MLP \\
    \hline
    Accuracy & 0.985 & 0.980 & 0.994 \\
    Precision & 0.992 & 0.970 & 0.993 \\
    Recall & 0.879 & 0.859 & 0.958 \\
    F1-Score & 0.927 & 0.906 & 0.975 \\
    \hline
  \end{tabular}
\end{table*}
\begin{figure*}[!t] 
\begin{minipage}[t]{0.32\textwidth}
\includegraphics[width=\textwidth]{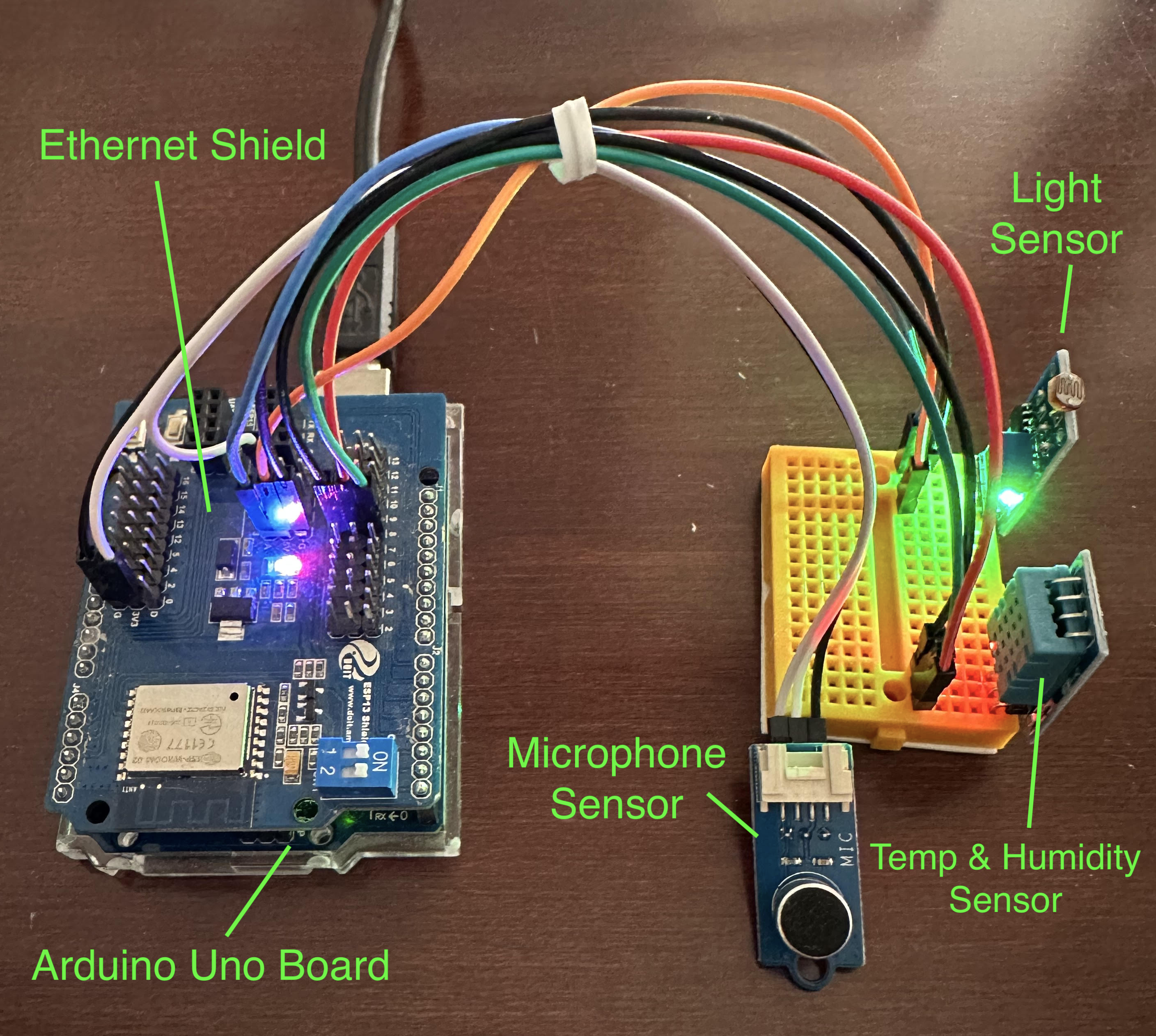}
\caption{IoT Node Implementation}
\label{fig:circuit} 
\end{minipage} \hfill
\begin{minipage}[t]{0.33\textwidth}
\raisebox{0.2\height}{\includegraphics[width=\textwidth]{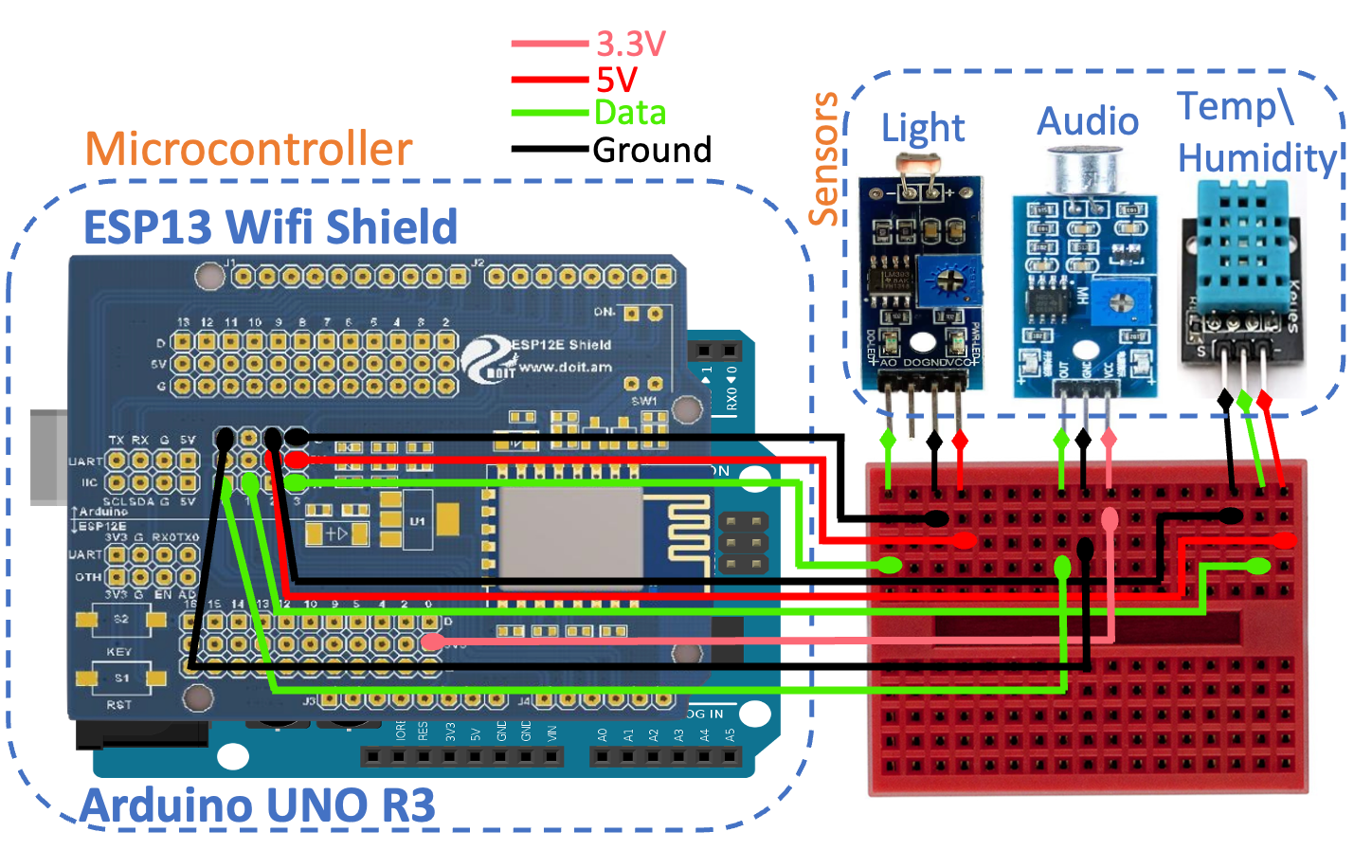}} 
\caption{Schematic Diagram}
\label{fig:schematic} 
\end{minipage} \hfill
\begin{minipage}[t]{0.32\textwidth}
\raisebox{0.1\height}{\includegraphics[width=\textwidth]{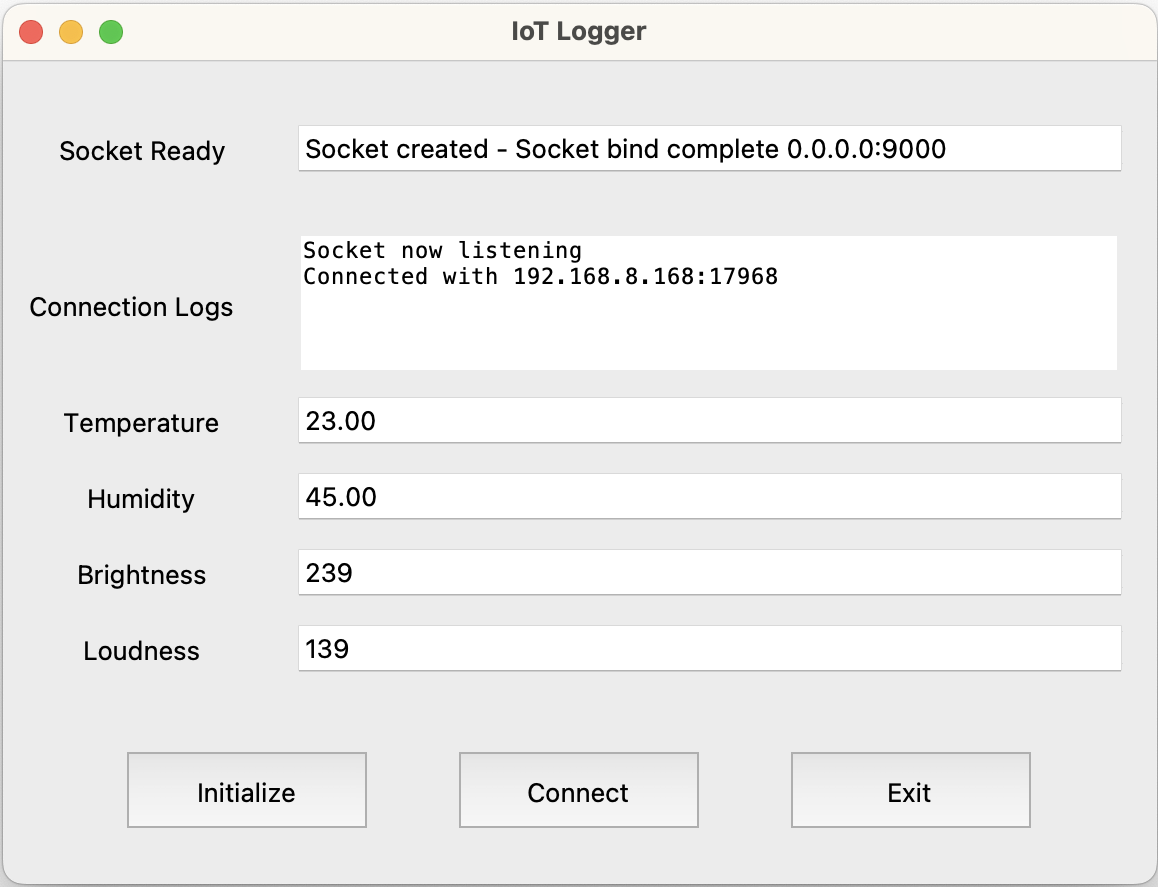}}
\caption{Software}
\label{fig:software} 
\end{minipage} \hfill
\end{figure*}
Table \ref{table:network-instrusion-accuracy} reports the accuracy for the same three classifiers used with the data anomaly models when applied to the network intrusion dataset. Notice that MLP produced the best results across all the metrics, proving our design choice of using it with tabular data that is large in size. When we compare this result with data anomaly we see that RF produced the best results mostly in all cases. This result can be attributed to the size of the IoTID20 dataset which is larger than the other two datasets. We note that the choice of which machine-learning classifier not only impacts the accuracy but also affects important measures relevant to system efficiency such as model size, model loading time and classification time. These measures are particularly important due to the distributed nature of our system architecture. Hence, we provide a complete system evaluation for the three datasets when utilizing the three classifiers in Section \ref{sec:evaluation}.
\subsection{IoT Node Implementation}
IoT nodes contain sensors and actuators to interact with the IoT environment. In addition, a networking module must be present through which the node can send and receive data and commands. We describe in this section an IoT node that we built to cover these requirements. The node represents the data anomaly in the AnoML-IoT scenario described in Section \ref{sec:design}. In this scenario, data from four sensors must be captured and mirrored to the DT namely temperature, humidity, loudness (i.e. microphone) and light sensors. Note that we did not implement the air quality sensor sense we noticed that the values of air quality in the dataset are not changing. By looking at Figure \ref{fig:circuit} we see that to add Wi-Fi capability to the Arduino board, we stacked in Ethernet shield (model: ESP13) on top of it. Afterwards, 5v, 3.3v, ground and analog connections of the sensors were connected to the ESP13 shield by means of jumper wires and a small breadboard. The schematic diagram for these connections is shown in Figure \ref{fig:schematic}. Notice from the figure that the light sensor (model: Photoresistor LDR Light Sensor Module) and temperature and humidity sensors (model: DHT11) needed to pull 5v current from the board. Whereas, the audio sensor pulled 3.3v current from the board to minimise the noise. The three sensor boards produced an analog output that was connected to the Arduino to be read over Wi-Fi. As shown in Figure \ref{fig:software}, we adapted the python software and implementation from hackster.io \cite{imp-software} and changed it to display our four sensor values. Once the python software is connected to the node it can poll the sensor values periodically and push them to the DT via Ditto APIs.

%% file: Evaluation.tex
\section{Performance Evaluation}
\label{sec:evaluation}
The overall performance of the architecture is impacted by how well the individual components perform as well as the overhead resulting from integrating those components. We begin in sections \ref{subsec: data-anomaly-performance-eval} and \ref{subsec: network-intrusion-performance-eval} by gauging the performance of the individual machine-learning models. After that, in Section \ref{subsec: arch-performance-eval} we measure the performance of the overall architecture when all the components are put together.
\subsection{Data Anomaly Performance Evaluation}
\label{subsec: data-anomaly-performance-eval}
\begin{table*}[!htb]
  \centering
  \caption{Time Performance for Data Anomaly Classifiers Based on Random Forest (RF), Support Vector Machines (SVM) and Multilayer Perceptron (MLP) ML Algorithms}
  \label{table:time-performance-data-anomaly}
  \begin{tabular}{lcccccccc}
    \hline
     & \multicolumn{4}{c}{AnoML-IoT} & \multicolumn{4}{c}{DS2OS} \\
     \cline{2-9}
     & Size & Loading (ms) & Fitting (ms) & Classification (ms) & Size & Loading (ms) & Fitting (S) & Classification (ms) \\
    \hline
    RF & 2.4 MB & 688.02 & 515.60 & 10.26 & 3.4 MB & 28.84 & 51.99 & 6.60 \\
    SVM & 41 KB & 1.16 & 126.12 & 0.34 & 20.60 MB & 25.19 & 284.58 & 2.76 \\
    MLP & 17 KB & 5.96 & 2,698.22 & 0.14 & 146 KB & 6.30 & 56.71 & 0.819 \\
    \hline
  \end{tabular}
\end{table*}
The choice of the dataset in terms of its size as well as the type of ML classifier utilized will have a significant impact on the performance of the architecture. The impact of these choices on the performance of the data anomaly model is shown in Table \ref{table:time-performance-data-anomaly}. We see from the table that the AnoML-IoT produced smaller model sizes compared to the DS2OS for all classifiers. This is normal since it is smaller in size with 6K records compared to 350K in the case of the DS2OS dataset. Consequently, the bigger size of the DS2OS models resulted a higher loading times compared to AnomML-IoT. Model loading time is an important metric as we can not assume that the model will always be readily available in memory when needed. Therefore, we cover the performance of both loaded models and unloaded models for all scenarios when we measure the overall architecture performance in Section \ref{subsec: arch-performance-eval}. Beside the loading time, we report the fitting time for the two datasets across the three machine learning models. We see that the fitting time is very high for the larger DS2OS dataset. We also see that the machine-learning model producing the highest fitting time for AnoML-IoT is MLP whereas the highest fitting time for the DS2OS is produced by SVM. This unpredictable performance for the fitting time is insignificant as the fitting process will take place offline. Finally, we see that time results were consistent for the classification time across the three machine-learning classifiers with RF having the highest classification time and MLP having the lowest. Classification time is very low (\textless 1ms) in the case of MLP making it the most suitable for use in the distributed architecture.
\subsection{Network Intrusion Performance Evaluation}
\label{subsec: network-intrusion-performance-eval}
\begin{table*}[!htb]
  \centering
  \caption{Time Performance for Network Intrusion Classifiers Based on Random Forest (RF), Support Vector Machines (SVM) and Multilayer Perceptron (MLP) ML Algorithms}
  \label{table:time-performance-network-intrusion}
  \begin{tabular}{lcccc}
    \hline
     & \multicolumn{4}{c}{IoTID20} \\
     \cline{2-5}
     & Size & Loading (ms) & Fitting (S) & Classification (ms) \\
    \hline
    RF & 26 MB & 69.83 & 44.77 & 5.92 \\
    SVM & 18.3 MB & 13.85 & 1489.21 & 14.27 \\
    MLP & 35 KB & 2.48 & 128.00 & 0.67 \\
    \hline
  \end{tabular}
\end{table*}
\begin{figure*}[!htb]
    \centering
    \includegraphics[width=0.95\textwidth]{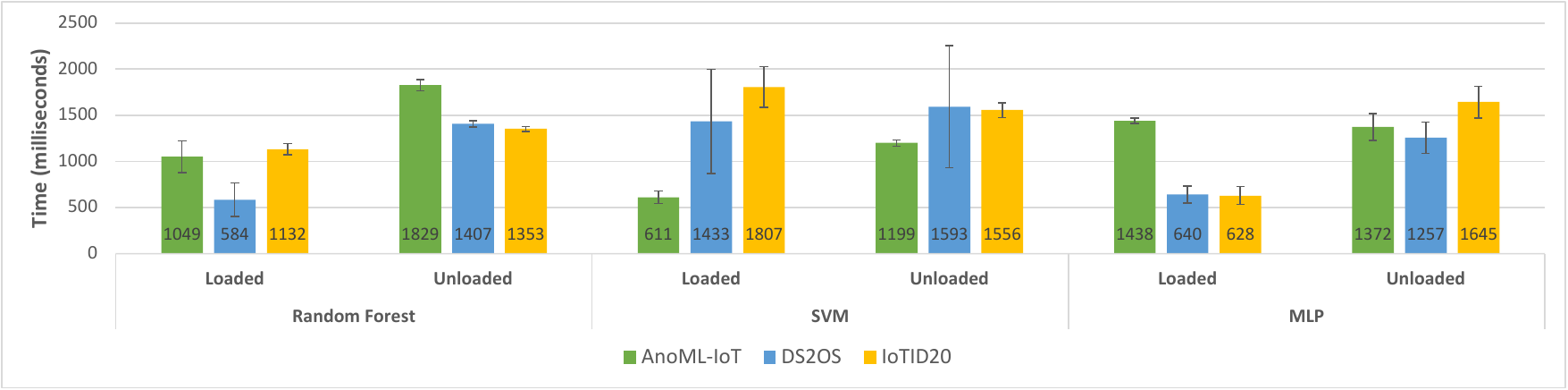}
    \caption{Time Elapsed Between DT State Change Until Receiving the Classification Result for Different Combinations of Datasets/ML Classifiers/Memory Loading Status}
    \label{fig:rtt-experiment}
\end{figure*}
We report the model size and time performance for the network intrusion model that is based on the IoTID20 dataset in Table \ref{table:network-instrusion-accuracy}. Notice that both RF and SVM created large model sizes in the order of MBs leading to higher loading times. Hence, these models must be carefully introduced to the architecture to ensure that models are pre-loaded so as to avoid impacting the performance. In terms of fitting time, the results where in harmony with the results achieved with the DS2OS dataset. The lowest fitting time was achieved by RF and the highest by SVM. Also, for classification time, MLP produced the lowest results making it the most suitable for use in a distributed setting.
\subsection{Architectures Performance Evaluation}
\label{subsec: arch-performance-eval}
Finally, we conducted an experiment to measure the performance of the architecture when different possibilities of individual components (i.e. datasets and classifiers) are utilized. Figure \ref{fig:rtt-experiment} reports the total time for detecting an anomaly when a change in the state of a DT is observed. Notice that we measure the total time at the fog server level since it has the administrative role in the environment in detecting malicious nodes and blocking them. More specifically, the time reported in the figure is the time elapsed between receiving a change in the state from the DT until the classification result is received from data anomaly or network intrusion container to the controller. Experiments cover both scenarios of a pre-loaded models in memory waiting for classification queries (Loaded) versus only loading a model when the request is received (unloaded). We report the average time from 5 runs for each experiment with standard error bars. Notice from the figure that in most of the cases, a loaded model achieved better time performance when compared to the corresponding case of an unloaded model. However, there are few exceptions (e.g. DS2OS/SVM and DS2OS/MLP scenarios) where the time is comparable. We attribute such performance variability to two reasons. First, Ditto' unstable performance leading to variability form one run to another. Second, Docker's \cite{docker} cashing behavior as docker pre-loads layers from previously loaded images to optimize the performance. Overall, we noticed from Figure \ref{fig:rtt-experiment} that when models are loaded the total time to detect an anomaly after observing a state changes is 500ms-600ms. This reasonable additional time is justified by the added security layer to detect and block malicious nodes in the environment.

%% file: Conclusion.tex
\section{Conclusion and Future Work}
\label{sec:conclusion}
This paper demonstrated the design, implementation, and evaluation of an architecture for compromised IoT node detection. The presented architecture has three important features. First, the concept of DT is utilized to only replicate relevant node information needed for malicious behavior detection. Second, classification is performed in the fog layer for low latency access to classifiers. Third, the architecture combines data anomaly and network intrusion detection methods for better coverage of malicious behavior detection leading to increased detection coverage. System evaluation of the architecture and the used classifiers demonestrates good accuracy as well as negligible overhead making it suitable for use in IoT solutions. In the future, we plan to vary system components related to the communication/application and study the impact on the performance. For example, connectivity using BLE or LoRa can be explored where BLE is suitable for body area networks and LoRa is suitable for smart agriculture. Also, we plan to test the system scalability by simulating calls from large number of IoT nodes and collecting larger volumes of data such as audio and video data. 